\documentclass{iau}
\usepackage{graphicx}
\usepackage{multirow}
\newcommand{\apj}{ApJ} 
\newcommand{\apjl}{ApJL} 
\newcommand{\mnras}{MNRAS}
\newcommand{\solphys}{SoPh}
\newcommand{\ssr}{SSR}
\newcommand{\jgr}{JGR}
\newcommand{\aap}{A\&A}
\newcommand{\araa}{Annual Review of Astronomy \& Astrophysics}

\begin{document}

\lefttitle{P.~Vemareddy}
\righttitle{Magnetic evolution of active regions}

\jnlPage{1}{7}
\jnlDoiYr{2024}
\doival{10.1017/xxxxx}

\aopheadtitle{Proceedings IAU Symposium 388}
\editors{N. Gopalswamy,  Olga E. Malandraki \&  Aline Vidotto, eds.}

\title{Magnetic evolution of active regions: formation and eruption of magnetic flux ropes}

\author{P.~Vemareddy}
\affiliation{Indian Institute of Astrophysics, II Block, Koramangala, Bengaluru-560 034, India \\
email: \email{vemareddy@iiap.res.in} }

\begin{abstract}
Magnetic flux ropes (FRs) are twisted structures appearing on the sun, predominantly in the magnetically concentrated regions. These structures appear as coronal features known as filaments or prominences in H$\alpha$ observations, and as sigmoids in X-ray, EUV observations. Using the continuous vector magnetic field observations from \textit{Helioseismic and Magnetic Imager} onboard \textit{Solar Dynamics Observatory}, we study the evolution of the magnetic fields in the active regions (ARs) to understand the conditions of twisted flux formation. While ARs emerge and evolve further, flux motions such as shearing and rotation are efficient mechanisms to form twisted flux ropes. Magnetic helicity quantifies the twisted magnetic fields and helicity injection through photosphere leads to its accumulation in the corona. Therefore, coronal helicity accumulation leads to twisted FR formation and its eruption. The magnetic helicity injection is seen to evolve distinctly in the regions of flux rope formation and eruption. The ARs that are associated with eruptive activity are observed with helicity injection predominantly with one sign over a period of a few days. The ARs that inject helicity with a changing sign are unlikely to form twisted FRs because coronal helicity during the period of one sign of injected helicity gets cancelled by the opposite sign of injection in the later period. As a result, the coronal field reconfigures from shared to potential structure. For a given AR, the upper limit of helicity that could cause a CME eruption is not yet understood, which can be the subject of future studies of ARs. Magnetic reconnection plays a crucial role in both the initiation and driving of FR eruptions after their formation. Data-driven simulations of the AR evolution provide more insights on the flux rope formation and its onset of eruption.
\end{abstract}


\maketitle

\section{Introduction}
\label{Intro}
Coronal mass ejections (CMEs) are large-scale magnetized plasma structures emanating from the solar atmosphere. CMEs occur more frequently and are most intense around the solar maximum. Fast CMEs drive interplanetary shocks, especially their propagation toward Earth, which was established to be the cause of the most severe geomagnetic disturbances on Earth (e.g.,  \citealt{Gosling1991_GeoMag_act,webb2000}). Geo-magnetic storms are disturbances in Earth's magnetosphere and can have a significant impact on both ground- and space-based technological systems. Therefore, it is of scientific and technological interest to understand the complete picture of CMEs, such as their origin from the source regions, evolution, and sun-to-earth propagation \citep{Webb2012_lrsp}.

In white-light observations, the CMEs are seen with a three-part structure morphology; viz. the leading edge, cavity, and core. These features of the CMEs are attributed to the compressed plasma in front of a flux rope (FR), a cavity, and a bright filament or prominence surrounding the cavity \citep{illing1985}. In this description, the magnetic configuration of the CME is a FR with a helical field that is wound around the central axis. A statistical study by \citet{Vourlidas2013} suggested that at least 40\% of CMEs observed by space-borne instruments have a clear FR structure. The in situ counterparts of the CMEs are known as Interplanetary Coronal Mass Ejections (ICMEs) which comprise magnetic clouds (MCs) of large-scale, organized magnetic structures \citep{burlaga1981}. The MCs are characterized by a smoothly rotating field of enhanced field strength from the background solar wind, low proton temperature, and low proton $\beta$. This characterization supports the FR topology to the in situ magnetic field and the MC is interpreted to be part of large scale bent FR extending from the Sun into interplanetary space \citep{burlaga1991}. 

\begin{figure*}
\centering
\includegraphics[width=0.99\textwidth,clip=]{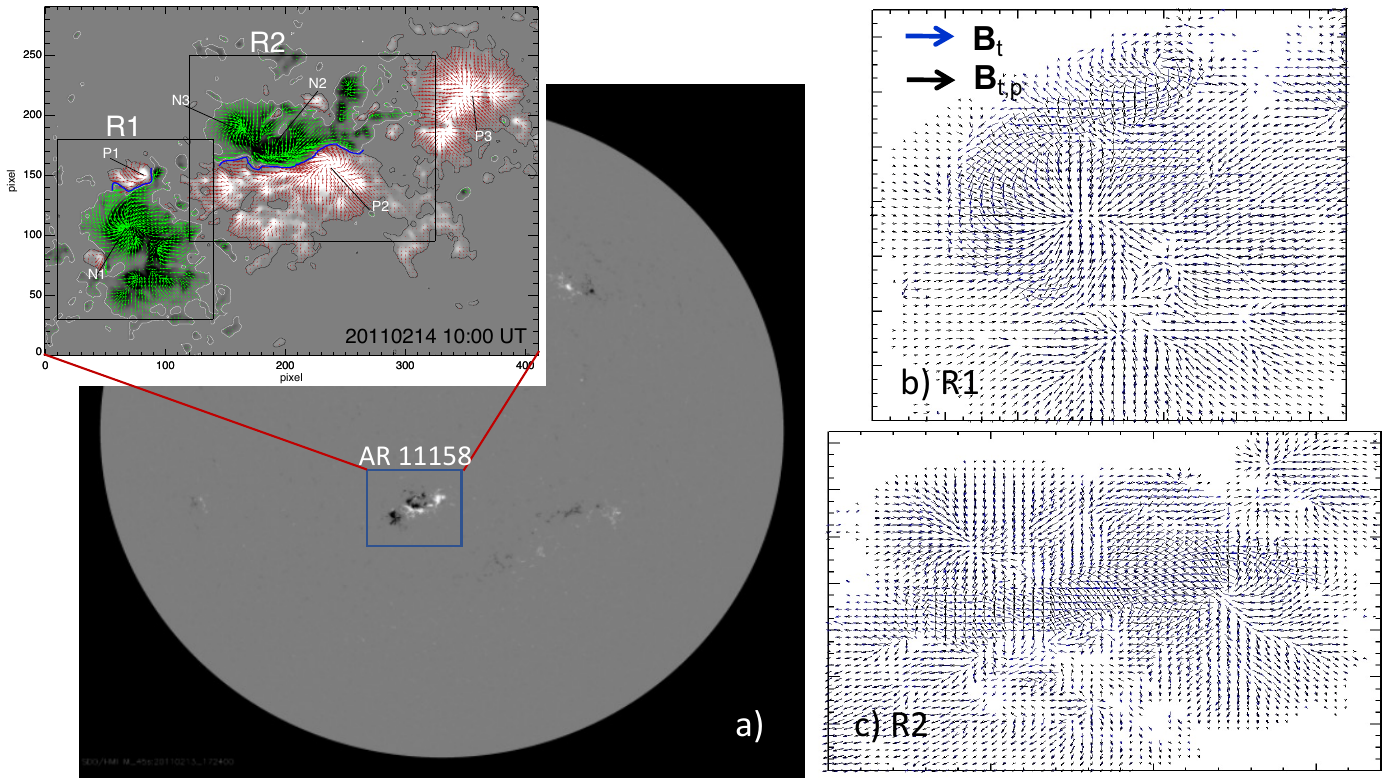}
\caption{A typical distribution of non-potential magnetic fields in an AR. a) HMI full-disk observation of a line-of-sight magnetic field. Inset is the vector magnetic field map of AR 11158 at 10:00 UT on February 14, 2011. The horizontal field vectors in red (green) are overplotted on the vertical component of the magnetic field with $\pm$150G contours. Major sunspot polarities are marked as P/N* within the rectangular regions of interest R1 and R2 (sub-regions) for further reference. The blue solid curves represent the strongly sheared (with shear angle greater than 45$^o$) PILs. The axis units are in pixels of 0.5 arcsec. b-c) Horizontal magnetic fields (blue) in sub-regions R1 and R2, respectively. Horizontal magnetic field vectors (black) derived from potential field extrapolation are overplotted to show the non-potentiality of the observed fields with respect to the flux motions in the AR. }
\label{fig1}
\end{figure*}

On the other hand, because the CMEs are linked to eruptions of solar features such as prominences, filaments, and sigmoids, the origin of the twisted FR must be in the source regions of these features. Therefore, to understand the origins of the CMEs on the Sun, it is very fundamental to study the physical mechanisms of the formation and eruption of FRs in the source regions. 

Magnetic fields play a fundamental role in the structure and dynamics of the solar corona. Driven by continuous plasma motions at foot points, the coronal field is stressed to build up electric current, and magnetic free energy. Such magnetic fields explode as CMEs/flares to release the free energy into thermal and kinetic energies \citep{shibata2011}. Sheared and twisted magnetic structures represent a non-potential magnetic field with free magnetic energy. Using the unprecedented, uninterrupted imaging observations from the Solar Dynamics Observatory (SDO; \citealt{pesnell2012}), in this proceeding paper, we study the magnetic evolution of the active regions (ARs) to comprehend the formation of sheared/twisted magnetic field and its subsequent eruption.  Having connection to the CME flux rope, we refer both sheared and twisted magnetic fields as twisted magnetic structures or FRs, although they are different from the modeling point of view. This paper is organized as follows: In section~\ref{MagEvol}, as an exemplary study, we showed the formation of twisted magnetic structure during the evolution of AR 11158. The characteristic evolution of ARs that is very likely to form the sheared/twisted magnetic field is studied by evaluating the helicity injection and its coronal accumulation in section~\ref{HelInj}. The triggering mechanisms of the eruptions are outlined in Section~\ref{TrigMech}, followed by a brief summary of the paper in Section~\ref{Summ}.

\begin{figure*}
\centering
\includegraphics[width=.99\textwidth,clip=]{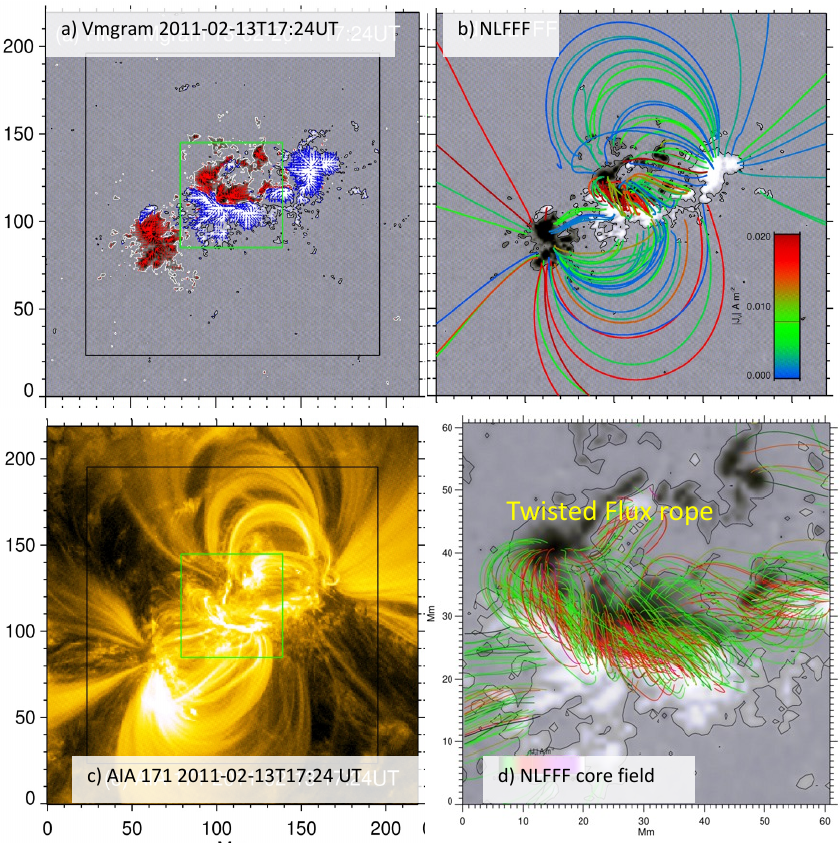}
\caption{ Magnetic field structure of AR 11158. a)  vector magnetogram used as the bottom boundary for the coronal field extrapolation, b) NLFFF magnetic structure rendered on the vertical magnetic field map, c) image of the AR corona in AIA 171~\AA~as proxy of coronal magnetic fields for comparison, d) magnetic structure at the core of the AR (green rectangular box) indicating highly twisted flux system, appearing as bright plasma emission in 171~\AA. }
\label{fig2}
\end{figure*}

\section{Magnetic evolution of the ARs and the twisted structures }
\label{MagEvol}
Magnetic field evolution is better studied with continuous, high-cadence observations of the full disk magnetic fields. Routine vector magnetic field observations (\texttt{hmi.sharp.cea\_720s} data product) are available from the Helioseismic and Magnetic Imager (HMI; \citealt{schou2012}) onboard SDO. As an example, we study the AR 11158, which has been highly flare-productive since its emergence on February 11, 2011 and has been the subject of several studies (e.g., \citealt{schrijver2011, vemareddy2012_sunspot_rot,sunx2012,Vemareddy2015_FluxEmer,Green2022_MagHel}). Its disk passage was well captured by multiple instruments, especially SDO. Fig~\ref{fig1}a shows the map of vector magnetic fields of the AR 11158 on February 14 at 10:00 UT. The major sunspot polarities are labelled (positive as P1, P2, P3 and negative as N1, N2, N3) and the regions of interest are enclosed in rectangular boxes (R1 and R2). During the evolution, the polarities underwent shear motions predominantly in sub-region R1 and rotational motions in sub-region R2. As a result, the horizontal field vectors are stressed in these regions. Such horizontal fields are mostly aligned parallel to the local polarity inversion line (PIL). The extent of stress is quantified by the shear angle, which is measured with respect to the horizontal field vectors of the potential field (PF). The potential field is the minimum energy stage of the magnetic field and is calculated with the $B_z$ component as a Neumann boundary condition. In Figure~\ref{fig1}(b-c), we show the horizontal vectors of both observed and potential magnetic fields in subregions R1 and R2. It can be evidently noticed that the sheared field mostly spread around the PIL. 

The buildup of magnetic non-potentiality during the AR evolution is monitored by parameters such as magnetic shear, force-free parameter ($\alpha_{av}$), vertical current ($J_z$), helicity injection rate ($dH/dt$), magnetic free- energy ($E_f$), etc. These parameters are found to have good correspondence with the rotating sunspot N1 in R1, as reported by \citet{vemareddy2012_sunspot_rot}. Moreover, they found that the free energy exhibits a step-down decrease at the onset of the flares from sub-region R2. 

\begin{figure*}
\centering
\includegraphics[width=.99\textwidth,clip=]{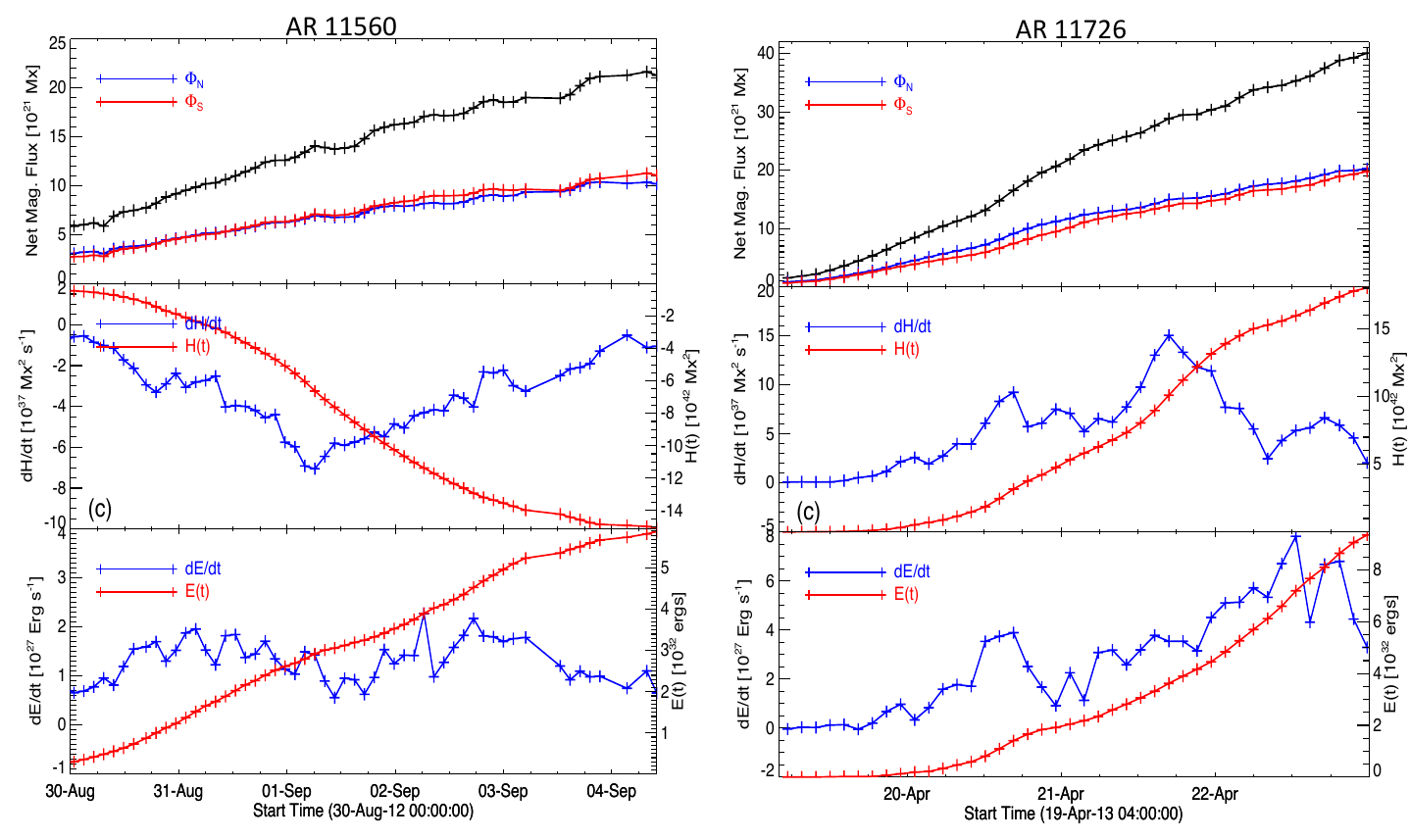}
\caption{Time evolution of magnetic flux, helicity, and energy fluxes in ARs 11560 (left) and 11726 (right). Time-integrated coronal helicity and energy (accumulated) quantities are plotted with the y-axis scale on the right of each plot. AR 11560 evolves with a predominant negative sign of $dH/dt$, whereas AR 11726 has a positive sign. Due to continuous injection of one sign of helicity flux, both these ARs have exhibited eruptive behavior. }
\label{fig3}
\end{figure*} 

Since regular measurements of coronal magnetic fields are not available, one extrapolates the observed photospheric vector field into the corona using the force-free approximation to the coronal field \citep{Wiegelmann2004}. In this model, the force-free parameter $\alpha$ varies spatially due to the presence of a sheared field, therefore known as non-linear force-free fields (NLFFFF). In Figure~\ref{fig2}a, we display the vector magnetogram used to construct NLFFF. Figure~\ref{fig2}b shows the rendered magnetic field of the NLFFF model, and is compared with the coronal plasma tracers observed in the AIA 171~\AA~image (Figure~\ref{fig2}c). From the comparison, we see that the NLFFF model reproduced structures similar to the coronal plasma tracers, and would be a promising coronal magnetic field model. In addition, the magnetic field at the core of the AR (blue rectangular box) is highly twisted, as displayed in Figure~\ref{fig2}d. Due to strong electric currents in the twisted field, the core field region is co-spatial with the bright plasma emission. The electric currents are non-neutralized in the regions of twisted flux \citep{Torok2014,Vemareddy2019_DegEle}. 

Thus, during the evolution of the AR, the flux motions drive the coronal field to non-potential form and form twisted FR. Shear and rotational motions are efficient energy storage mechanisms to account for CMEs and flares. The FRs thus formed are subject to erupt once they attain the critical limits of energy and helicity \citep{zhangmei2005}. The eruption of the FRs is the only way to release the excess free magnetic energy, which is being converted to kinetic and thermal energy. An X2.2 flare associated with large CME and several C and M class flares occurred from subregion R2 of AR 11158 where shear motions are predominant. Some ARs were reported to launch successive CMEs due to the continuous process of FR formation and its eruption under a constant driver of shear and converging flux motions \citep{Vemareddy2017_SuccHomol}.

\begin{figure*}
\centering
\includegraphics[width=.99\textwidth,clip=]{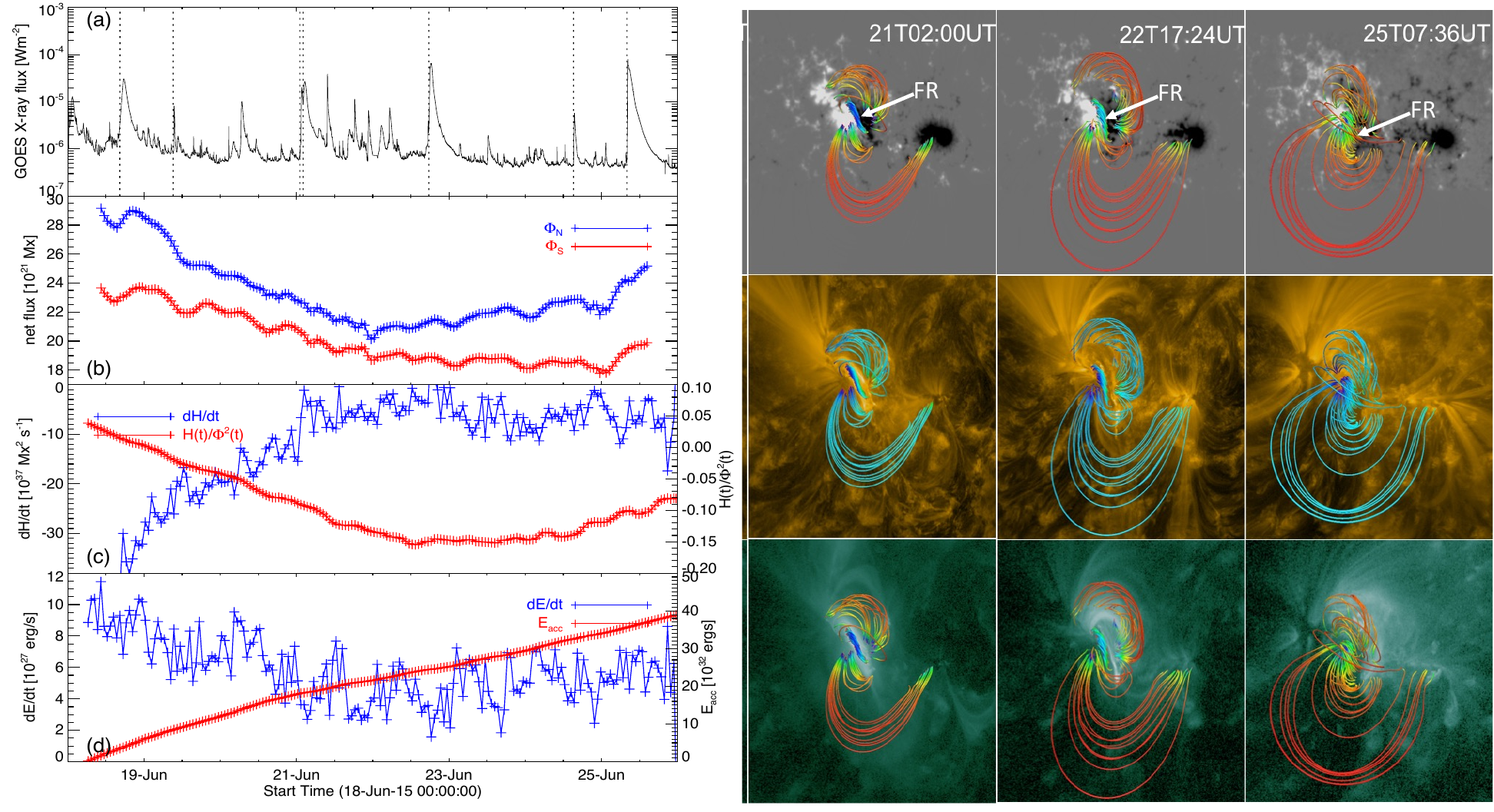}
\caption{Magnetic evolution in successively erupting AR 12371. {\bf left column panels:} time profiles of disk-integrated soft x-ray flux (a), net magnetic flux (b), $dH/dt$ (c) and $dE/dt$ (d). {\bf right column panels:} Rendered NLFFF magnetic structure on $B_z$ map and AIA 171\AA~ and 94\AA~maps at different times. While helicity flux continuously accumulates with a predominant sign, the coronal magnetic fields store energy to form twisted flux along the sheared PIL.  }
\label{fig4}
\end{figure*} 

\section{Characteristic helicity injection for the helicity accumulation in the AR corona}
\label{HelInj}

Magnetic helicity is a measure of magnetic twist and shear which can be used to quantify the complexity of the AR magnetic field. Due to conserved nature, magnetic helicity transfers from one volume to another. Especially in the case of the sun, its transport through photospheric surface in the open coronal volume is  estimated by the mathematical expression derived by \citet{Berger1984},

\begin{equation}
{{\left. \frac{dH}{dt} \right|}_{S}}=2\int\limits_{S}{\left( {{\mathbf{A}}_{P}}\bullet {{\mathbf{B}}_{t}} \right){{\text{V}}_n}dS}-2\int\limits_{S}{\left( {{\mathbf{A}}_{P}}\bullet {{\mathbf{V}}_{t}} \right){{\text{B}}_{n}}dS}\\
\label{eq_dhdt}
\end{equation} 
with the help of velocity ($\mathbf{V}$) and magnetic field $\mathbf{B}$ observations. This equation essentially quantifies the complexity generated by the flux motions, including emerging (first term) and shear motions (second term), as the ARs emerge and evolve further. The velocity $\mathbf{V}$ is derived by tracking the flux motions from time sequence magnetic field observations \citep{schuck2005, schuck2008}. Using time sequence line-of-sight magnetic field observations of the AR, \citet{Chae2001_ObsDetMagHel} for the first time used this equation to compute the helicity injection rate ($dH/dt$), followed by several other researchers (e.g., \citealt{Nidos2003_MagHelBud, Labonte2007_SurveyMagHel, Park2010_ProdFlareMagHel, vemareddy2012_hinj}) to understand connection of helicity injection with the occurrence of large flares. Similarly, the energy injection is computed as 

\begin{equation}
{{\left. \frac{dE}{dt} \right|}_{S}}=\frac{1}{4\pi }\int\limits_{S}{B_{t}^{2}{{V}_{n}}dS-\frac{1}{4\pi }}\int\limits_{S}{\left( {{\mathbf{B}}_{t}}\bullet {{\mathbf{V}}_{t}} \right){{B}_{n}}dS}
\label{eq_dedt}
\end{equation}

Given the profiles of energy and helicity injections rates, one integrates over time to estimate their coronal budgets. Unlike the statistical studies for large flare productivity \citep{Labonte2007_SurveyMagHel, Park2010_ProdFlareMagHel}, helicity injection and its coronal accumulation have implications to the FR formation followed by its eruption and it was conjectured that the CMEs are the only way to remove excess coronal helicity \citep{zhangmei2005}. For the formation of the twisted flux during the AR evolution, the $dH/dt$ has to be predominantly one sign over a long period. In order to understand the fundamental role of magnetic helicity in large scale eruptions, we need to characterise the $dH/dt$ profile in different ARs producing eruptive or non-eruptive flares. For this purpose, we studied a few emerging ARs along with their generated activity.

\begin{figure*}
\centering
\includegraphics[width=.99\textwidth,clip=]{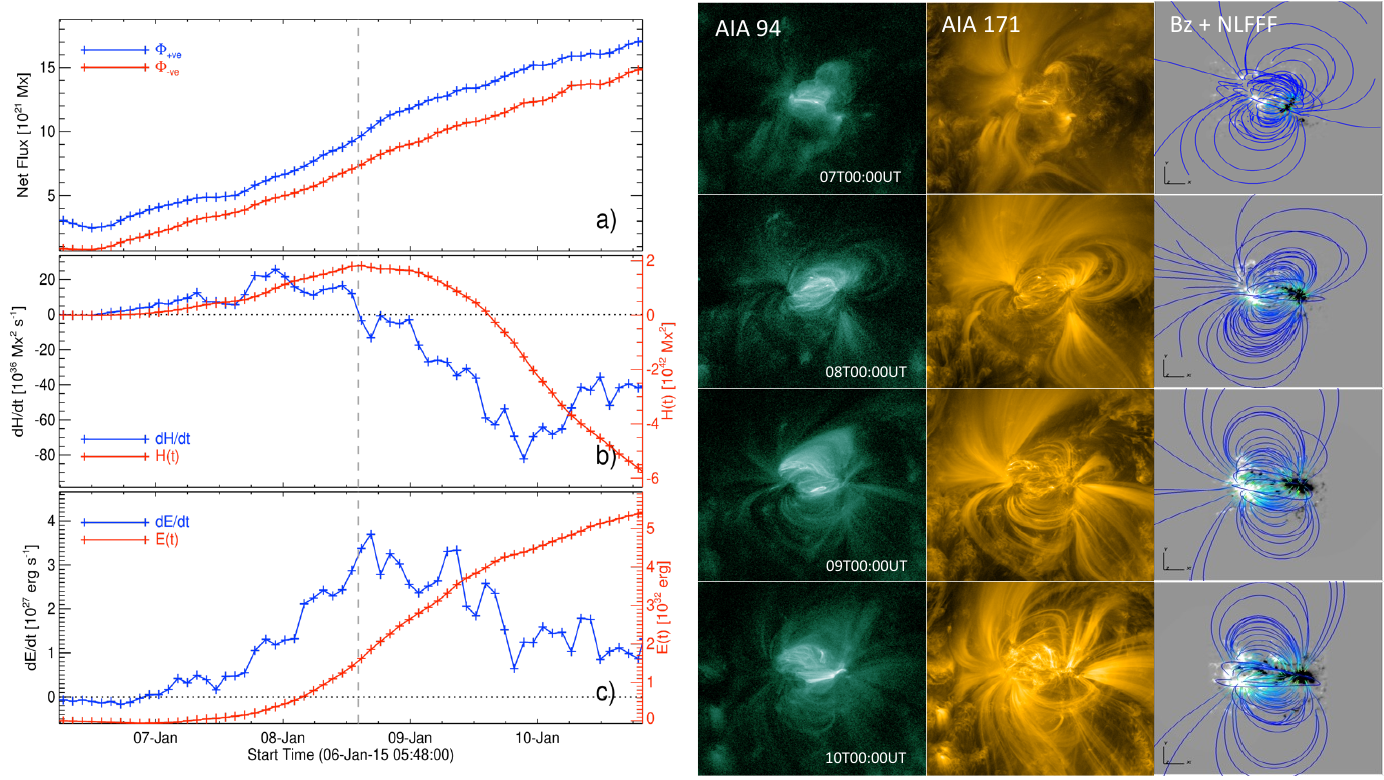}
\caption{Magnetic evolution of the AR 12257 without exhibiting eruptive behavior. Left column panels: time evolution of net magnetic flux, the helicity injection rate ($dH/dt$), and the energy injection rate. Accumulated helicity, $H(t)$, and energy, $E(t)$, are plotted with their scales on the right-side axis. Vertical-dashed line marks the time (2015 January 08T14:00 UT) when $dH/dt$ changes sign from positive to negative. Right column panels: Comparison of NLFFF magnetic structure with plasma tracers in EUV images in AIA 94, 171~\AA~images at different epochs of the AR evolution.  }
\label{fig5}
\end{figure*}

Figure~\ref{fig3} shows magnetic evolution in emerging ARs 11560 and 11726 during their disk transit within $\pm$40 longitude. While magnetic polarities emerge and evolve, the $dH/dt$ is predominant with a sign, so the coronal helicity is additive in time. Note the difference in injection rate in these two ARs because the flux motions (both emerging and shear) are different in different ARs. As a result, these ARs exhibit eruptive behaviour at some point in time. The time profiles of $dH/dt$ are examined in several such ARs. In Figure~\ref{fig4}, we plot the magnetic evolution in AR 12371 from where successive CME eruptions were launched over the course of 7 days of evolution. The $dH/dt$ profile is negative all the time, then there could be instances when the accumulated helicity exceeds the critical value, and as a result, the FR eruption occurs intermittently as CMEs. Although the AR is in the decay phase, the core polarities are in shear, and converging motions, which form twisted flux along the PIL. The pre-eruptive magnetic structure of NLFFF model reveals the existence of a twisted FR along the PIL of shearing polarities \citep{Vemareddy2018}.

There are also ARs that have different evolutions of magnetic fluxes. In Figure~\ref{fig5}, we plot the magnetic evolution of AR 12257, which did not exhibit eruptive activity except for C-class flaring events. In this AR, the $dH/dt$ changes sign from positive to negative values, and as a consequence, the helicity accumulation is not additive. In this case, the twisted flux may not form, but the magnetic structure may transform from one chiral (handedness) form to another. In the right-hand panels of Figure~\ref{fig5}, the rendered NLFFF magnetic structure at different epochs presents a sheared magnetic configuration without a twisted FR at the core. And the sheared field during positive $dH/dt$ reforms into a potential field during negative $dH/dt$. Such a detailed study of helicity injection in ARs, supported by the coronal field model, unambiguously supports the theoretical idea of the conservation property of magnetic helicity being the cause of CME eruptions. More cases of opposite helicity injecting ARs were reported in \citet{Vemareddy2022_NatHel}.

\section{Triggering Mechanism of solar eruptions}
\label{TrigMech}
When the AR has reached a state of stored energy configuration through continuous flux motions, its sudden release requires a suitable triggering mechanism. There have been several ideas explored for explaining the onset mechanism of the eruptions \citep{forbes2006, moore2006,Perenti2014_SolarProm}. All these models are based on a pre-eruptive configuration, either as a sheared arcade or a twisted FR. In the sheared arcade based models, the erupting feature is assumed to be composed of the sheared and twisted field in the core of the AR, and then the internal/external runaway tether-cutting reconnection forms the FR and its subsequent eruption \citep{moore2001, antiochos1999, amari2003a}. On the other hand, FR based models invoke a twisted FR that could lose equilibrium due to ideal MHD instability such as kink or torus instability \citep{torok2004}. Because of the wide variety of dynamic processes occurring in complex ARs, however, it may be difficult to identify a particular triggering mechanism responsible for an observed eruption.

\begin{figure*}
\centering
\includegraphics[width=.97\textwidth,clip=]{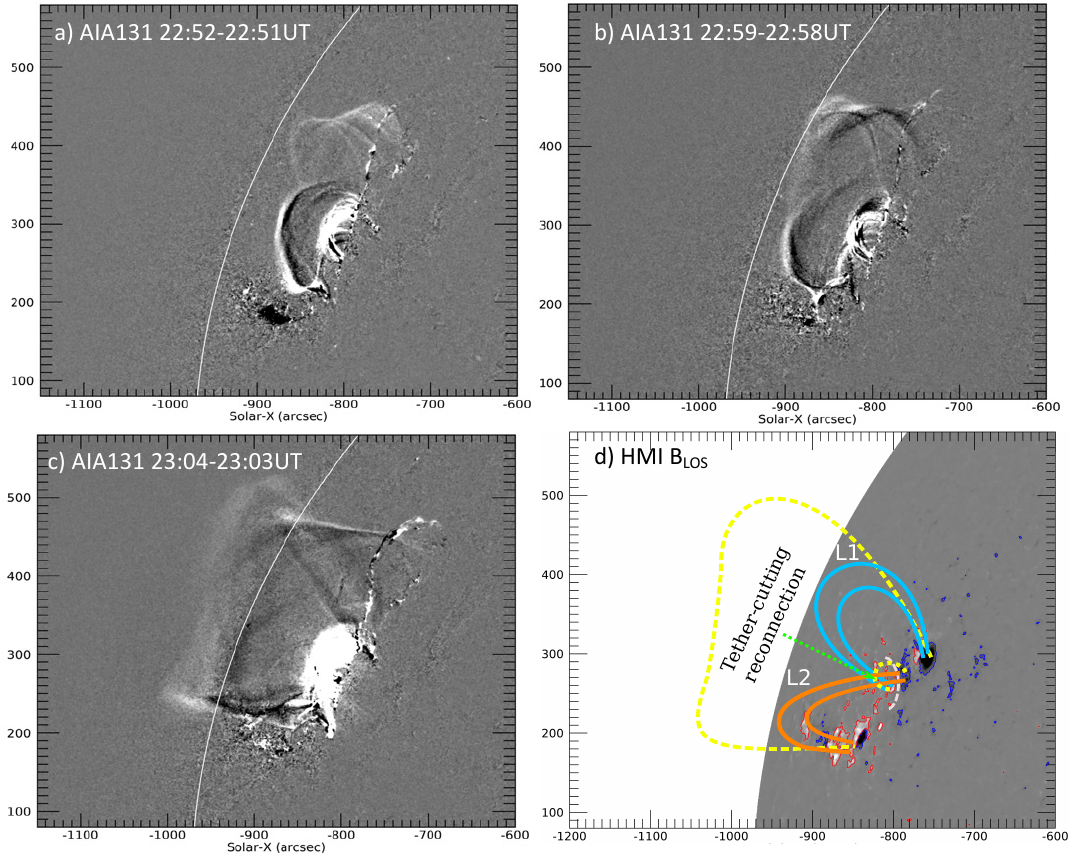}
\caption{Evidence for tether-cutting reconnection as the initiation of the eruption. a-c) difference images of AIA 131~\AA~during the onset of the eruption. Two nearby loop structures, L1 and L2, first rise up, then reconnect at a coronal cross-point to form a single larger loop structure with flare loops below. b) A schematic illustration of the possible magnetic structure undergoing the tether-cutting reconnection.}
\label{fig6}
\end{figure*}

Observational evidence for tether cutting reconnection forming the FR and triggering its eruption were reported in several studies (e.g., \citealt{vemareddy2012_FilErup, LiuChang2013_EvidenceTether, Xue2017_FormFluxRope_Tether, ChenHechao2018_Witnessing_TethCutt, Vemareddy2022_ErupEUVHot}. Typically in this model, two adjacent opposite sheared loops cross each other over the PIL as shown in Figure~\ref{fig6}(a-b). In top-view observations (e.g., \citealt{Vemareddy2017_SuccHomol}) they appear as two lobes of a sigmoid. When subject to a converging motion toward the PIL, these loops reconnect to form upward-rising twisted field lines, seen as continuous EUV structure in observations (Figure~\ref{fig6}c). The schematic in Figure~\ref{fig6}d better depicts this scenario, linking the foot points of the loops with the magnetic field observations. The reconnection also induces the flare emission underneath the FR and acts as a driving mechanism to further accelerate the FR. Similarly, there were observational studies that claimed direct evidence of ideal kink instability with the flux system underlying the prominence/filament body (e.g., \citealt{rust1996, Green2007, gilbert2007, Vemareddy2017_PromEru}). After the FR reaches a certain height where the magnetic field decrease is critically steep, the torus instability plays a role in driving the eruption further \citep{Vasantharaju2019_FindCritDecayInd, aulanier2010}. In both types of these mechanisms, reconnection plays a crucial role by transforming the stabilizing upper arcade into the twisted field surrounding the original core field.

\section{Summary}
\label{Summ}
We study the magnetic evolution of the ARs which is the central subject of the eruptive behavior of the sun. While ARs emerge and evolve further, flux motions constantly drive the coronal magnetic field to a stored magnetic energy configuration. Shearing and rotational motions of the fluxes are especially efficient mechanisms to form twisted magnetic structures in the form of FRs. FR formation is a consequence of helicity accumulation in the corona \citep{zhangmei2005}, and for that, helicity injection from the surface has to be predominantly one sign over a period of a few days. The ARs that inject helicity with changing signs are unlikely to form twisted FRs because coronal helicity during the period of one sign of injected helicity gets cancelled by the opposite sign of injection in the later period. As a result, the coronal field reconfigures itself from sheared to potential magnetic structure. For a given AR, the upper limit of helicity that could cause a CME eruption is not yet understood, which can be the subject of further studies of ARs.

The FR or sheared magnetic configurations are basic ingredients in most of the eruption models. Magnetic reconnection plays a crucial role in both the initiation and driving of FR eruptions. Data-driven simulations of the AR evolution provide more insights to advance our understanding of the FR formation and its onset of eruption.

{\bf Acknowledgements:} SDO is a mission of NASA's Living With a Star Program; STEREO is the third mission of NASA's Solar-Terrestrial Probes program; and SOHO is a mission of international cooperation between the ESA and NASA. The author acknowledges the collaborations with Prof. Jie Zhang, Dr. Pascal Dem\"oulin, and Dr. Nat Gopalswamy on some of the research results discussed in this paper. The author is thankful to the referee for reading and commenting on this draft.

\bibliographystyle{iaulike.bst}

\end{document}